# Method Article – Title Page

| | |
|---|---|
| **Title** | BSF-skeleton: A Template for Parallelization of Iterative Numerical Algorithms on Cluster Computing Systems |
| **Authors** | Leonid B. Sokolinsky |
| **Affiliations** | South Ural State University,<br>76, Lenin prospekt, Chelyabinsk, Russian Federation. 454080 |
| **Corresponding Author's email address** | leonid.sokolinsky@susu.ru |
| **Keywords** | • parallel computation model<br>• C++<br>• MPI<br>• master/slave framework<br>• higher-order function<br>• Map/Reduce<br>• scalability boundary prediction |
| **Direct Submission or Co-Submission** | *Co-Submission* DOI: 10.1016/j.jpdc.2020.12.009 |

**ABSTRACT**

This article describes a method for creating applications for cluster computing systems using the parallel BSF-skeleton based on the original BSF (Bulk Synchronous Farm) model of parallel computations developed by the author earlier. This model uses the master/slave paradigm. The main advantage of the BSF model is that it allows to estimate the scalability of a parallel algorithm before its implementation. Another important feature of the BSF model is the representation of problem data in the form of lists that greatly simplifies the logic of building applications. The BSF-skeleton is designed for creating parallel programs in C++ using the MPI library. The scope of the BSF-skeleton is iterative numerical algorithms of high computational complexity. The BSF-skeleton has the following distinctive features.
- The BSF-skeleton completely encapsulates all aspects that are associated with parallelizing a program.
- The BSF-skeleton allows error-free compilation at all stages of application development.
- The BSF-skeleton supports OpenMP programming model and workflows.

**SPECIFICATIONS TABLE**

| | |
|---|---|
| **Subject Area** | Computer Science |
| **More specific subject area** | Parallel programming |
| **Method name** | BSF parallel skeleton |
| **Name and reference of original method** | Bulk Synchronous Farm: parallel computation model<br><br>L.B. Sokolinsky, BSF: a parallel computation model for scalability estimation of iterative numerical algorithms on cluster computing systems, Journal of Parallel and Distributed Computing, 149 (2021): pp. 193–206. https://doi.org/10.1016/j.jpdc.2020.12.009. |
| **Resource availability** | Source code is freely available at https://github.com/leonid-sokolinsky/BSF-skeleton |

## Method details

A parallel skeleton is a programming construct, which abstracts a pattern of parallel computation and interaction [1]. The BSF-skeleton extends *farm* skeleton based on the master/slave paradigm. The *farm* skeleton and the master/slave paradigm are discussed in a large number of papers (see, for example, [2–5]). A distinctive feature of the BSF-skeleton is that it combines *farm*, *map*, and *reduce* algorithmic skeletons. The theoretical basis of the BSF-skeleton is the BSF (Bulk Synchronous Farm) model of parallel computations [6]. The BSF-skeleton uses the master/worker (master/slave) paradigm to organize interaction between MPI processes (see Fig. 1). This means that worker processes can only exchange messages with the master process. To use the BSF-skeleton, you must represent your algorithm in the form of operations on lists using the higher-order functions *Map* and *Reduce* [7]. The higher-order function $Map(f, A)$ applies the function $f$ to each element of list $A = [a_1, \ldots, a_n]$ converting it to the list $B = [f(a_1), \ldots, f(a_n)]$. The higher-order function $Reduce(\oplus, B)$ taking an associative binary operation $\oplus$ and a list $B = [b_1, \ldots, b_n]$ as parameters calculates the element $b = b_1 \oplus \ldots \oplus b_n$. One should use Algorithm 1 as a template. Let us comment on Algorithm 1. The variable $i$ denotes the iteration number; $x^{(0)}$ is an initial approximation; $x^{(i)}$ is the $i$-th approximation (the approximation can be a number, a vector, or any other data structure); $A$ is the list of elements of a certain set $\mathbb{A}$, which represents the source data of the problem; $F_x: \mathbb{A} \to \mathbb{B}$ is a parameterized user function (the parameter $x$ is the current approximation) that maps the set $\mathbb{A}$ to a set $\mathbb{B}$; $B$ is a list of elements of the set $\mathbb{B}$ calculated by applying the function $F_x$ to each element of the list $A$; $\oplus$ is an binary associative operation on the set $\mathbb{B}$.

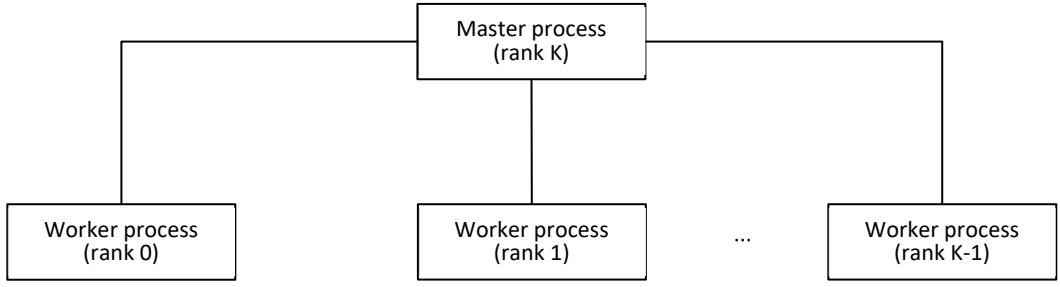

**Fig. 1.** Interaction of $K+1$ MPI processes in the BSF-skeleton.

Step 1 reads input data of the problem and an initial approximation. Step 2 assigns the zero value to the iteration counter $i$. Step 3 calculates the list $B$ by applying the higher-order function $Map(F_{x^{(i)}}, A)$. Step 4 assigns the result of the higher-order function $Redice(\oplus, B)$ to the intermediate variable $s \in \mathbb{B}$. Step 5 invokes the user function *Compute* that calculates the next approximation $x^{(i+1)}$ taking two parameters: the current approximation $x^{(i)}$ and the result $s$ of the higher-order function *Reduce*. Step 6 increases the iteration counter $i$ by one. Step 7 checks a termination criteria by invoking the Boolean user function *StopCond*, which takes two parameters: the new approximation $x^{(i)}$ and the previous approximation $x^{(i+1)}$. If *StopCond* returns true, the algorithm outputs $x^{(i)}$ as an approximate problem solution and stops working. Otherwise, the control is passed to Step 3 starting the next iteration.

**Algorithm 1.** Generic BSF-algorithm template.

1: **input** $A, x^{(0)}$
2: $i := 0$
3: $B := Map(F_{x^{(i)}}, A)$
4: $s := Compute(\oplus, B)$
5: $x^{(i+1)} := Compute(x^{(i)}, s)$
6: $i := i + 1$
7: **if** $StopCond(x^{(i)}, x^{(i-1)})$ **goto** 9
8: **goto** 3
9: **output** $x^{(i)}$
10: **stop**

The BSF-skeleton automatically parallelizes Algorithm 1 by splitting the list $A$ into $K$ sublists of equal length ($\pm 1$):

$$A = A_0 + \cdots + A_{K-1},$$

where $K$ is the number of worker processes and $+\!\!+\!$ denotes the operation of list concatenation. This uses the parallelization scheme shown in Fig. 2.

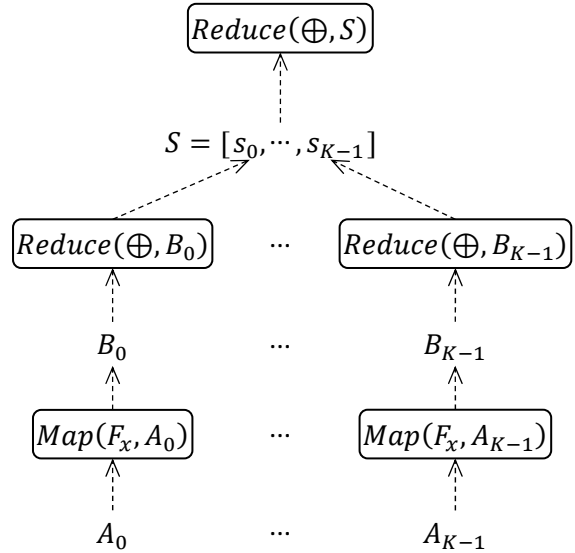

**Fig. 2.** BSF-skeleton parallelization schema.

The result is the parallel Algorithm 2. It includes $K + 1$ parallel processes: one master process and $K$ worker processes. In Step 2, the master process sends the current approximation $x^{(i)}$ to all worker processes. After that, every $j$-th worker process independently applies higher-order function *Map* and *Reduce* to its sublist (the steps 3 and 4). In the steps 3 and 4, the master process is idle. In Step 5, all worker processes send the partial foldings $s_0, \ldots, s_{K-1}$ to the master process. In the steps 6-9, the master process performs the following actions: executes the higher-order function *Reduce* over the list of partial foldings $[s_0, \ldots, s_{K-1}]$; invokes the user function *Compute* that calculates the next approximation; checks the termination criteria by using the Boolean user function *StopCond* and assigns its result to the Boolean variable *exit*. In the steps 6-9, the worker processes are idle. In Step 10, the master process sends the *exit* value to all worker processes. If the *exit* value is false, the master process and worker processes go to the next iteration, otherwise the master processes outputs the result and the computation stops. Note that, in the Steps 2 and 10, all processes perform the implicit global synchronization.

**Algorithm 2.** BSF-skeleton parallelization template.

| Master | j-th worker (j=0,...,K-1) |
|---|---|
| 1: **input** $x^{(0)}$; $i \coloneqq 0$ | 1: **input** $A_j$ |
| 2: $SendToAllWorkers(x^{(i)})$ | 2: $RecvFromMaster(x^{(i)})$ |
| 3: | 3: $B_j \coloneqq Map(F_{x^{(i)}}, A_j)$ |
| 4: | 4: $s_j \coloneqq Reduce(\oplus, B_j)$ |
| 5: $RecvFromWorkers(s_0, \ldots, s_{K-1})$ | 5: $SendToMaster(s_j)$ |
| 6: $s \coloneqq Reduce(\oplus, [s_0, \ldots, s_{K-1}])$ | 6: |
| 7: $x^{(i+1)} \coloneqq Compute(x^{(i)}, s)$ | 7: |
| 8: $i \coloneqq i + 1$ | 8: |
| 9: $exit \coloneqq StopCond(x^{(i)}, x^{(i-1)})$ | 9: |
| 10: $SendToAllWorkers(exit)$ | 10: $RecvFromMaster(exit)$ |
| 11: **if** $exit$ **goto** 2 | 11: **if** $exit$ **goto** 2 |
| 12: **output** $x^{(i)}$ | 12: |
| 13: **stop** | 13: **stop** |

## Source code structure of BSF-skeleton

The BSF-skeleton is a compilable but not executable set of files. This set is divided into two groups: files with the "*BSF*" prefix contain problem-independent code and are not subject to changes by the user; files with the "*Problem*" prefix are intended for filling in problem-dependent parts of the program by the user. Descriptions of all source code files are given in Table 1.

**Table 1.** Source code files of the BSF-skeleton.

| File | Description |
|---|---|
| *Problem-independent code* | |
| *BSF-Code.cpp* | Implementations of the *main* function and all problem-independent functions |
| *BSF-Data.h* | Problem-independent variables and data structures |
| *BSF-Forwards.h* | Declarations of the problem-independent functions |
| *BSF-Include.h* | The inclusion of problem-independent libraries |
| *BSF-SkeletonVariables.h* | Definitions of the skeleton variables (see Section "Skeleton variables") |
| *BSF-ProblemFunctions.h* | Declarations of the problem-dependent BSF functions (see Section "Predefined problem-dependent BSF functions (prefix PC_bsf_)") |
| *BSF-Types.h* | Definitions of problem-independent types |
| *Problem-dependent code* | |
| *Problem-bsfCode.cpp* | Implementations of the problem-dependent BSF functions (see Section "Predefined problem-dependent BSF functions (prefix PC_bsf_)") |
| *Problem-bsfParameters.h* | BSF-skeleton parameters (see Section "BSF-skeleton parameters") |
| *Problem-bsfTypes.h* | Predefined BSF types (see Section "**Ошибка! Источник ссылки не найден.**") |
| *Problem-Data.h* | Problem-dependent variables and data structures |
| *Problem-Forwards.h* | Declarations of the problem-dependent functions |
| *Problem-Include.h* | Inclusion of problem-dependent libraries |
| *Problem-Parameters.h* | Parameters of the problem |
| *Problem-Types.h* | Problem types |

The dependency graph of the source code files by the directive *#include* is shown in Fig. 3. The gray rectangles indicate the code files that do not allow changes. The rectangles with striped shading indicate the code files containing predefined declarations that must be defined (filled in) by the user. The white rectangles indicate the code files that should be fully implemented by the user.

## BSF-skeleton parameters

The BSF-skeleton parameters are declared as macroses in the file *Problem-bsfParameters.h*. They are used in the *BSF-Code.cpp* and should be set by the user. All these parameters are presented in Table 2.

## Predefined problem-depended BSF types

The predefined problem-depended BSF types are declared as data structures in the file *Problem-bsfTypes.h*. They are used in the *BSF-Code.cpp* and should be set by the user. All these types are presented in Table 3.

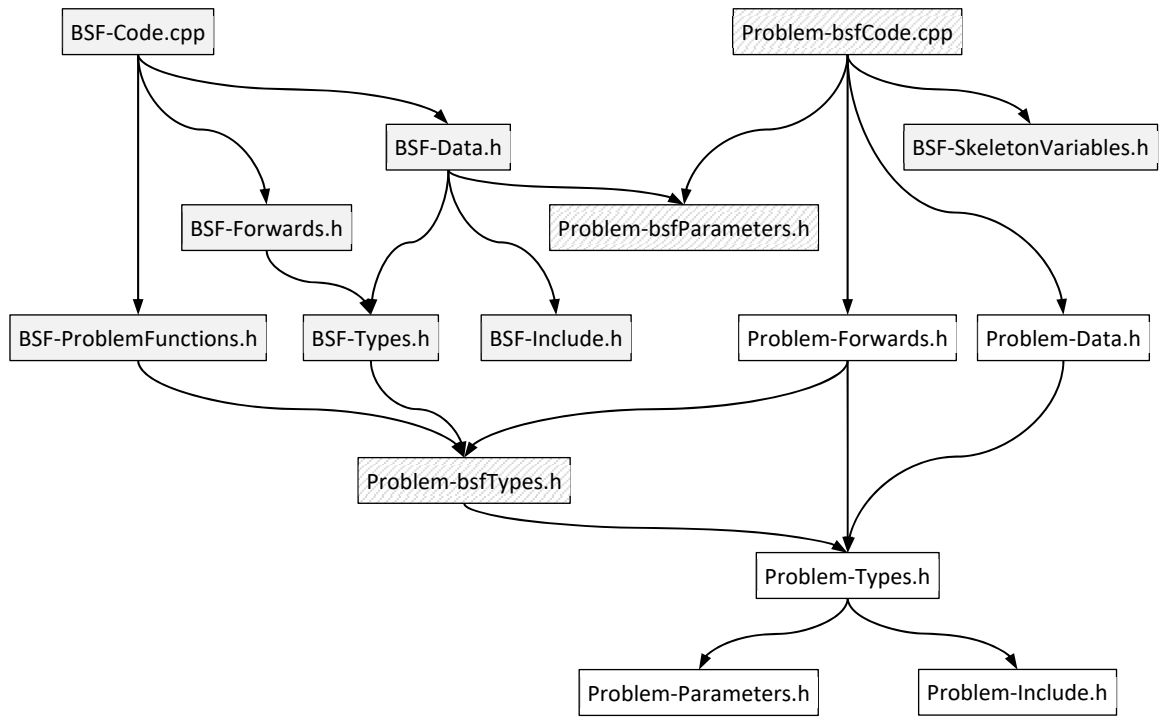

**Fig. 3.** Dependency graph of the source code files by the directive #include.

**Table 2.** Predefined problem-dependent parameters.

| ID | Description | Default value |
|---|---|---|
| *PP_BSF_MAX_MPI_SIZE* | Defines the maximum possible number of MPI processes (the result returned by the function *MPI_Comm_size* cannot exceed this number). | 500 |
| *PP_BSF_PRECISION* | Sets the decimal precision to be used to format floating-point values on output operations. | 4 |
| *PP_BSF_ITER_OUTPUT* | If this macros is defined, at the end of each *k*-th iteration, the master process will invocate the predefined BSF function *PC_bsf_IterOutput* that outputs intermediate results. The number *k* is defined by the macros *PP_BSF_TRACE_COUNT*. | #undef |
| *PP_BSF_TRACE_COUNT* | Defines the number *k* mentioned in the description of the macros *PP_BSF_ITER_OUTPUT*. | 1 |
| *PP_BSF_MAX_JOB_CASE* | Defines the maximum number of activities (jobs) in workflow minus 1. See Section "Workflow support". | 0 |
| *PP_BSF_OMP* | If this macros is defined, the worker processes use *#pragma omp parallel for* to perform the higher-order function *Map*. | #undef |
| *PP_BSF_NUM_THREADS* | If this macros is defined, *omp parallel for* uses the specified number of threads to perform the higher-order function *Map*. If this macros is not defined, *omp parallel for* uses the maximum possible number of threads. | #undef |

**Table 3.** Predefined BSF types (file *Problem-bsfTypes.h*).

| Type ID | Data type | Description | Mandatory to fill in |
|---|---|---|---|
| *PT_bsf_parameter_T* | struct | Defines the structure (set of data elements) that is transferred by the master process to all the worker processes and includes the order parameters (usually the current approximation). | Yes |
| *PT_bsf_mapElem_T* | struct | Defines the record that represents an element in the map-list (list *A* in Algorithm 1). | Yes |
| *PT_bsf_reduceElem_T* | struct | Defines the record that represents an element in the reduce-list[1] (list *B* in Algorithm 1). | Yes |
| *PT_bsf_reduceElem_T_1, PT_bsf_reduceElem_T_2, PT_bsf_reduceElem_T_3* | struct | Alternative types of the reduce-list elements that are used to organize the workflow (see Section "Workflow support"). | No |

## Extended reduce-list

The BSF-skeleton appends to each element of the reduce-list the additional integer field called *reduceCounter*. This extended reduce-list is presented by the pointer *BD_extendedReduceList* declared in the *BSF-Data.h*. When performing the *Reduce* function (see *BC_ProcessExtendedReduceList* in Section "Key problem-independent functions (prefix *BC_*)"), the elements that have this field equal to zero are ignored. For elements where *reduceCounter* is not zero, the values of the *reduceCounter* are added together. By default, the function *BC_WorkerMap* (see Section "Key problem-independent functions (prefix *BC_*)") sets the *reduceCounter* to 1. The user can set the value of this field to 0 by setting the parameter *success* of the function *PC_bsf_MapF* to 0.

---

[1] *Reduce-list* is the list being the second parameter of the higher-order function *Reduce*.

## Skeleton variables

The skeleton variables are declared in the file *BSF-SkeletonVariables.h*. The user can exploit these variables for the sake of debugging, tracing, and non-standard implementing (see, for example, Section "Using Map without Reduce"). The user should not change the values of these variables. All skeleton variables are presented in Table 4.

**Table 4.** Skeleton variables (file *BSF-SkeletonVariables.h*).

| Skeleton variable | Type | Description |
|---|---|---|
| BSF_sv_addressOffset | int | Contains the number of the first element of the map-sublist appointed to the current worker process. |
| BSF_sv_iterCounter | int | Contains the number of iterations performed so far. |
| BSF_sv_jobCase | int | Contains the number of the current activity (job) in workflow (see Section "Workflow support"). |
| BSF_sv_mpiMaster | int | Contains the rank (number) of the master MPI process. |
| BSF_sv_mpiRank | int | Contains the rank (number) of current MPI process. |
| BSF_sv_numberInSublist | int | This variable contains the relative number of the element in the map-sublist that the function *Map* is currently applied to. |
| BSF_sv_numOfWorkers | int | Contains the total number of the worker processes. |
| BSF_sv_parameter | PT_bsf_parameter_T | Structure that contains the order parameters. |
| BSF_sv_sublistLength | int | Contains the length of the map-sublist appointed to a worker process. |

## Functions

The skeleton functions are divided into two groups:
1) problem-independent functions with the prefix *BC_* that have implemented in the file *BSF-Code.cpp*;
2) problem-dependent functions (predefined BSF functions) with the prefix PC_bsf_ that have declared in the file Problem-Code.cpp.

The user cannot change the headers and bodies of the functions with the prefix *BC_*. The user also cannot change function headers with the prefix *PC_bsf_* but must write an implementation of these functions. The body of a predefined BSF function cannot include calls of problem-independent functions with the prefix *BC_*. The hierarchy of the key function calls is presented in Fig. 4.

## Key problem-independent functions (prefix *BC_*)

The implementations of all problem-independent functions can be found in the file *BSF-Code.cpp*. Descriptions of some key problem-independent functions are presented in Table 5.

## Predefined problem-dependent BSF functions (prefix PC_bsf_)

This section contains detailed descriptions of the predefined problem-dependent BSF functions with the prefix *PC_bsf_* declared in *Problem-bsfCode.cpp*. The user must implement all these functions. An instruction is presented in Section "Step-by-step instruction". An example is presented in Section "Example of using the BSF-skeleton".

### PC_bsf_CopyParameter

Copies all order parameters from the in-structure to the out-structure. The order parameters are declared in the predefined problem-depended BSF type *PT_bsf_parameter_T* (see Section "Ошибка! Источник ссылки не найден.").

*Syntax*
```
void PC_bsf_CopyParameter(
        PT_bsf_parameter_T parameterIn,
        PT_bsf_parameter_T* parameterOutP
);
```
*In parameters*
parameterIn
    The structure from which parameters are copied.
*Out parameters*
parameterOutP
    The pointer to the structure to which parameters are copied.

### PC_bsf_Init

Initializes the problem-depended variables and data structures defined in *Problem-Data.h*.
*Syntax*
```
void PC_bsf_Init(
        bool* success
);
```
*Out parameters*
*success
    Must be set to *false* if the initialization failed. The default value is *true*.

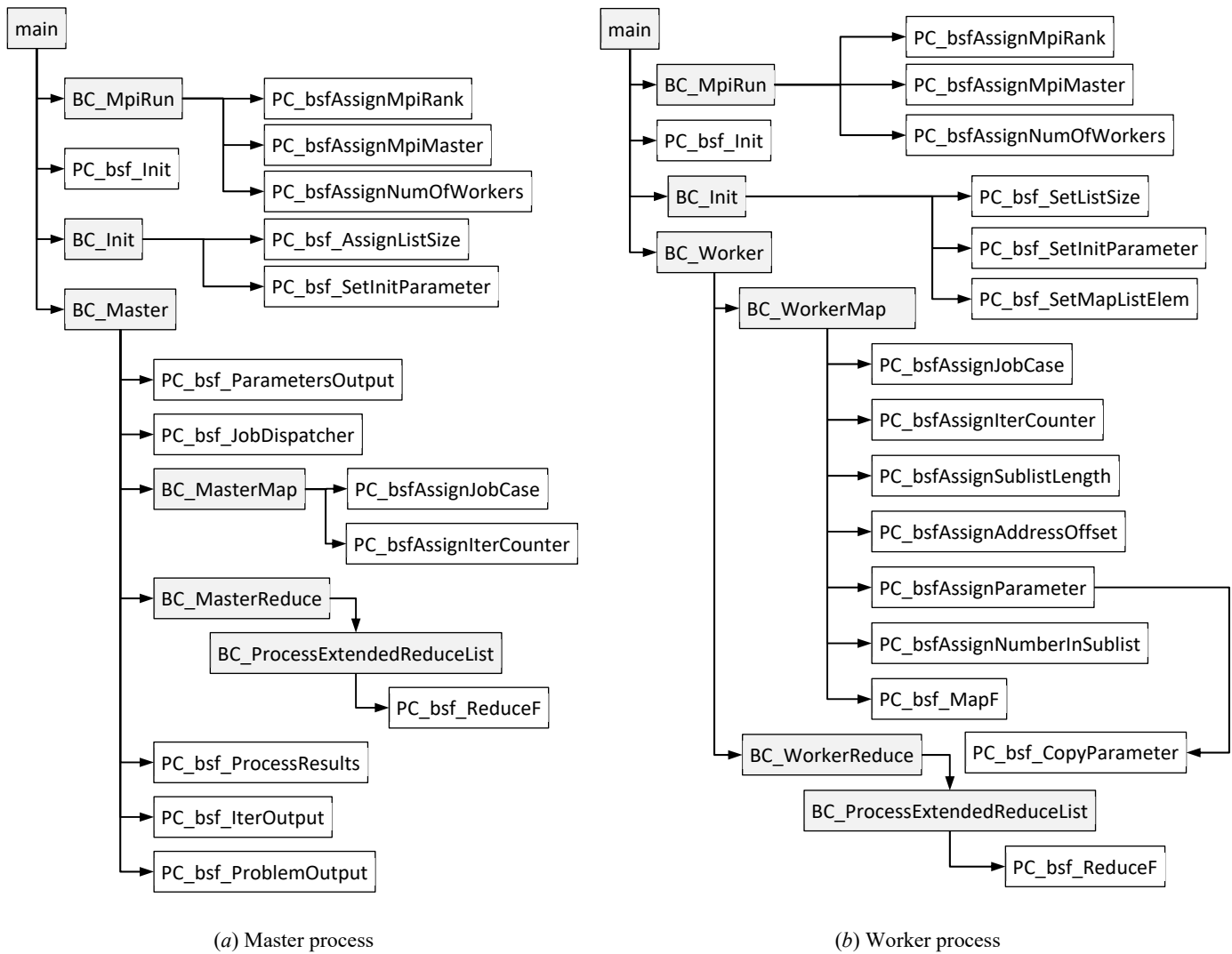

Fig. 4. Hierarchy of the key function calls.

**PC_bsf_IterOutput**

    Outputs intermediate results of the current iteration.

*Syntax*
```
void PC_bsf_IterOutput(
        PT_bsf_reduceElem_T* reduceResult,
        int reduceCounter,
        PT_bsf_parameter_T parameter,
        double elapsedTime,
        int nextJob
);
void PC_bsf_IterOutput_1(
        PT_bsf_reduceElem_T_1* reduceResult,
        int reduceCounter,
        PT_bsf_parameter_T parameter,
        double elapsedTime,
        int nextJob
);
 PC_bsf_IterOutput_2(
        PT_bsf_reduceElem_T_2* reduceResult,
        int reduceCounter,
        PT_bsf_parameter_T parameter,
        double elapsedTime,
        int nextJob
);
void PC_bsf_IterOutput_3(
        PT_bsf_reduceElem_T_3* reduceResult,
        int reduceCounter,
        PT_bsf_parameter_T parameter,
        double elapsedTime,
        int nextJob
);
```

**Table 5.** Key problem-independent functions (file *BSF-Code.cpp*).

| Function | Description |
|---|---|
| BC_Init | Performs the memory allocation and the initialization of the skeleton data structures and variables. |
| BC_Master | The head function of the master process. |
| BC_MasterMap | Forms an order and sends it to the worker processes to perform the *Map* function in the current iteration. |
| BC_MasterReduce | Receives the results produced by the worker processes, collects them in a list, and performs the function *Reduce* on this list. |
| BC_MpiRun | Executes the MPI initialization. After it, the number of worker processes is accessible by the skeleton variable *BSF_sv_numOfWorkers*; total number of MPI processes (*MPI_Comm_size*) is equal to ($BSF_sv_numOfWorkers + 1$); the rank of a MPI process (*MPI_Comm_rank*) is accessible by the skeleton variable *BSF_sv_mpiRank*; the rank of the master MPI process is accessible by the skeleton variable *BSF_sv_mpiMaster* (is equal to *MPI_Comm_size*-1). The MPI ranks of the worker processes have values from 0 to ($BSF_sv_numOfWorkers - 1$). The MPI rank of the worker process is equal to *BSF_sv_numOfWorkers*. |
| BC_ProcessExtendedReduceList | This function finds the first element in the extended reduce-list with the *reduceCounter* not equal to zero and adds to it all other elements that have the *reduceCounter* not equal to zero. For pairwise addition of elements of the original reduce-list, the function *PC_bsf_ReduceF* is used. |
| BC_Worker | The head function of a worker process. |
| BC_WorkerMap | Receives the order from the master process, assigns the skeleton variables (see Section "Skeleton variables"), and applies the function *PC_bsf_MapF* to the appointed map-sublist to produce the corresponding part of the reduce-list. |
| BC_WorkerReduce | Sends to the master process the element that is the sum of all reduce-sublist elements. |

*In parameters*

reduceResult
    Pointer to the structure that contains the result of executing the *Reduce* function.
reduceCounter
    The number of summed (by $\oplus$) elements in the reduce-list. This number matches the number of extended reduce-list elements that have the value 1 in the field *reduceCounter* (see Section "Extended reduce-list").

*Remarks*

    The functions *PC_bsf_IterOutput_1*, *PC_bsf_IterOutput_2* and *PC_bsf_IterOutput_3* are used to organize a workflow (optional filling).

**PC_bsf_JobDispatcher**

    This function is used to organize the workflow (see Section "Workflow support") and is executed by the master process before starting each iteration. It implements a state machine that switches from one state to another. If you do not need the workflow support, then you should use the empty implementation of this function.

*Syntax*

```
void PC_bsf_JobDispatcher(
PT_bsf_parameter_T* parameter,
int* job,
bool* exit
);
```

*In/out parameters*

parameter
    The pointer to the structure containing the parameters of the next iteration. This structure may be also modified by the functions *PC_bsf_ProcessResults_1*, *PC_bsf_ProcessResults_2* and *PC_bsf_ProcessResults_3*.

*Out parameters*

*job
    This variable must be assigned the number of the next action (job).
*exit
    If the stop condition holds, then this variable must be assigned *true*. The default value is *false*.

*Remarks*

    Important: The use of the structure *BSF_sv_parameter* is not allowed in the implementation of this function.
    The function *PC_bsf_JobDispatcher* is invocated after the invocation of function *PC_bsf_ProcessResults_1*, *PC_bsf_ProcessResults_2* or *PC_bsf_ProcessResults_3*.

## PC_bsf_MapF

Implements the function that is applied to the map-list elements when performing the higher-order function *Map*. To implement the *PC_bsf_MapF* function, we can use the problem-dependent variables and data structures defined in the file *Problem-Data.h*, and the structure *BSF_sv_parameter* of the type *PT_bsf_parameter_T* defined in *Problem-bsfTypes.h*.

*Syntax*
```
void PC_bsf_MapF(
    PT_bsf_mapElem_T* mapElem,
    PT_bsf_reduceElem_T* reduceElem,
    int* success
);
void PC_bsf_MapF_1(
    PT_bsf_mapElem_T* mapElem,
    PT_bsf_reduceElem_T_1* reduceElem,
    int* success
);
void PC_bsf_MapF_2(
    PT_bsf_mapElem_T* mapElem,
    PT_bsf_reduceElem_T_2* reduceElem,
    int* success
);
void PC_bsf_MapF_3(
    PT_bsf_mapElem_T* mapElem,
    PT_bsf_reduceElem_T_3* reduceElem,
    int* success
);
```

*In parameters*

mapElem
: The pointer to the structure that is the current element of the map-list.

*Out parameters*

reduceElem
: The pointer to the structure that is the corresponding reduce-list element to be calculated.

*success
: Must be set to *false* if the corresponding reduce-list element must be ignored when the *Reduce* function will be executed. The default value is *true*.

*Remarks*

The functions *PC_bsf_MapF_1*, *PC_bsf_MapF_2* and *PC_bsf_MapF_3* are used to organize a workflow (optional filling).

## PC_bsf_ParametersOutput

Outputs parameters of the problem before starting the iterative process.

*Syntax*
```
void PC_bsf_ParametersOutput(
    PT_bsf_parameter_T parameter
);
```

*In parameters*

parameter
: The structure containing the parameters of the problem.

## PC_bsf_ProblemOutput

Outputs the results of solving the problem.

*Syntax*
```
void PC_bsf_ProblemOutput(
    PT_bsf_reduceElem_T* reduceResult,
    int reduceCounter,
    PT_bsf_parameter_T parameter,
    double t
);
void PC_bsf_ProblemOutput_1(
    PT_bsf_reduceElem_T_1* reduceResult,
    int reduceCounter,
    PT_bsf_parameter_T parameter,
    double t
);
void PC_bsf_ProblemOutput_2(
    PT_bsf_reduceElem_T_2* reduceResult,
    int reduceCounter,
    PT_bsf_parameter_T parameter,
    double t
);
void PC_bsf_ProblemOutput_3(
    PT_bsf_reduceElem_T_3* reduceResult,
    int reduceCounter,
    PT_bsf_parameter_T parameter,
    double t
);
```

*In parameters*

reduceResult
    The pointer to the structure that is the result of executing the higher-order function *Reduce*.
parameter
    The structure containing the parameters of the final iteration.

*Remarks*

    The functions *PC_bsf_ProblemOutput_1*, *PC_bsf_ProblemOutput_2* and *PC_bsf_ProblemOutput_3* are used to organize a workflow (optional filling).

## PC_bsf_ProcessResults

    Processes the results of the current iteration: computes the order parameters for the next iteration and checks the stop condition.

*Syntax*

```
void PC_bsf_ProcessResults(
    PT_bsf_reduceElem_T* reduceResult,
    int reduceCounter,
    PT_bsf_parameter_T* parameter,
    int* nextJob,
    bool* exit
);
void PC_bsf_ProcessResults_1(
    PT_bsf_reduceElem_T_1* reduceResult,
    int reduceCounter,
    PT_bsf_parameter_T* parameter,
    int* nextJob,
    bool* exit
);
void PC_bsf_ProcessResults_2(
    PT_bsf_reduceElem_T_2* reduceResult,
    int reduceCounter,
    PT_bsf_parameter_T* parameter,
    int* nextJob,
    bool* exit
);
void PC_bsf_ProcessResults_3(
    PT_bsf_reduceElem_T_3* reduceResult,
    int reduceCounter,
    PT_bsf_parameter_T* parameter,
    int* nextJob,
    bool* exit
);
```

*In parameters*

reduceResult
    The pointer to the structure that is the result of executing the higher-order function *Reduce*.
reduceCounter
    The number of summed (by $\oplus$) elements in the reduce-list. This number matches the number of extended reduce-list elements that have the value 1 in the field *reduceCounter* (see Section "Extended reduce-list").

*In/out parameters*

parameter
    The pointer to the structure containing the parameters of the current iteration. This structure must be modified by setting new values of the parameters for the next iteration.

*Out parameters*

*nextJob
    If a workflow is used (see Section "Workflow support"), then this variable must be assigned the number of the next action (job). Otherwise, this parameter is not used.
*exit
    If the stop condition holds, then this variable must be assigned *true*. The default value is *false*.

*Remarks*

    <span style="color:red">Important</span>: The use of the structure *BSF_sv_parameter* is not allowed in the implementations of these functions.
    The functions *PC_bsf_ProcessResults_1*, *PC_bsf_ProcessResults_2* and *PC_bsf_ProcessResults_3* are used to organize a workflow (optional filling).

## PC_bsf_ReduceF

    Implements the operation $z = x \oplus y$ (see Section "Method details").

*Syntax*

```
void PC_bsf_ReduceF(
    PT_bsf_reduceElem_T* x,
    PT_bsf_reduceElem_T* y,
    PT_bsf_reduceElem_T* z
);
void PC_bsf_ReduceF_1(
    PT_bsf_reduceElem_T_1* x,
    PT_bsf_reduceElem_T_1* y,
```

```
        PT_bsf_reduceElem_T_1* z
);
void PC_bsf_ReduceF_2(
        PT_bsf_reduceElem_T_2* x,
        PT_bsf_reduceElem_T_2* y,
        PT_bsf_reduceElem_T_2* z
);
void PC_bsf_ReduceF_3(
        PT_bsf_reduceElem_T_3* x,
        PT_bsf_reduceElem_T_3* y,
        PT_bsf_reduceElem_T_3* z
);
```
*In parameters*

x
    The pointer to the structure that presents the first term.

y
    The pointer to the structure that presents the second term.

*Out parameters*

z
    The pointer to the structure that presents the result of the operation.

*Remarks*

    The functions *PC_bsf_ReduceF_1*, *PC_bsf_ReduceF_2* and *PC_bsf_ReduceF_3* are used to organize a workflow (optional filling).

### PC_bsf_SetInitParameter

    Sets initial order parameters for the workers in the first iteration. These order parameters are declared in the predefined problem-depended BSF type *PT_bsf_parameter_T* (see Section "**Ошибка! Источник ссылки не найден.**").

*Syntax*
```
void PC_bsf_SetInitParameter(
        PT_bsf_parameter_T* parameter
);
```
*Out parameters*

parameter
    The pointer to the structure that the initial parameters should be assigned to.

### PC_bsf_SetListSize

    Sets the length of the list.

*Syntax*
```
void PC_bsf_SetListSize(
        int* listSize
);
```
*Out parameters*

*listSize
    Must be assigned a positive integer that specifies the length of the list.

*Remarks*

    The list size should be greater than or equal to the number of workers.

### PC_bsf_SetMapListElem

    Initializes the map-list element with the number *i*.

*Syntax*
```
void PC_bsf_SetMapListElem(
        PT_bsf_mapElem_T* elem,
        int i
);
```
*In parameters*

elem
    The pointer to the map-list element.

i
    The ordinal number of the specified element.

### Remarks

Important: The numbering of elements in the list begins from zero.

### PC_bsfAssignAddressOffset

    Assigns the number of the first element of the map-sublist to the skeleton variables *BSF_sv_addressOffset* (see Section "Skeleton variables").

*Syntax*
```
void PC_bsfAssignAddressOffset(int value);
```
*In parameters*

value

Non-negative integer value.

*Remarks*

Important: The user should not use this function.

## PC_bsfAssignIterCounter

Assigns the number of the first element of the map-sublist to the skeleton variables *BSF_sv_iterCounter* (see Section "Skeleton variables").

*Syntax*

```
void PC_bsfAssignIterCounter(int value);
```

*In parameters*

value

Non-negative integer value.

*Remarks*

Important: The user should not use this function.

## PC_bsfAssignJobCase

Assigns the number of the current activity (job) in workflow to the skeleton variables *BSF_sv_jobCase* (see Section "Skeleton variables").

*Syntax*

```
void PC_bsfAssignJobCase(int value);
```

*In parameters*

value

Non-negative integer value.

*Remarks*

Important: The user should not use this function.

## PC_bsfAssignMpiMaster

Assigns the rank of the master MPI process to the skeleton variables *BSF_sv_mpiMaster* (see Section "Skeleton variables").

*Syntax*

```
void PC_bsfAssignMpiMaster(int value);
```

*In parameters*

value

Non-negative integer value.

*Remarks*

Important: The user should not use this function.

## PC_bsfAssignMpiRank

Assigns the rank of current MPI process to the skeleton variables *BSF_sv_mpiRank* (see Section "Skeleton variables").

*Syntax*

```
void PC_bsfAssignMpiRank(int value);
```

*In parameters*

value

Non-negative integer value.

*Remarks*

Important: The user should not use this function.

## PC_bsfAssignNumberInSublist

Assigns the number of the current element in the map-sublist to the skeleton variables *BSF_sv_numberInSublist* (see Section "Skeleton variables").

*Syntax*

```
void PC_bsfAssignNumberInSublist(int value);
```

*In parameters*

value

Non-negative integer value.

*Remarks*

Important: The user should not use this function.

## PC_bsfAssignNumOfWorkers

Assigns the total number of the worker processes to the skeleton variables *BSF_sv_numOfWorkers* (see Section "Skeleton variables").

*Syntax*

```
void PC_bsfAssignNumOfWorkers(int value);
```

*In parameters*

value

Non-negative integer value.

*Remarks*

    Important: The user should not use this function.

**PC_bsfAssignParameter**

    Assigns the order parameters to the structure *BSF_sv_parameter* (see Section "Skeleton variables").

*Syntax*

```
void PC_bsfAssignParameter(PT_bsf_parameter_T parameter);
```

*In parameters*

parameter
    The structure from which the order parameters are taken.

*Remarks*

    Important: The user should not use this function.

**PC_bsfAssignSublistLength**

    Assigns the length of the current map-sublist to the skeleton variables *BSF_sv_sublistLength* (see Section "Skeleton variables").

*Syntax*

```
void PC_bsfAssignSublistLength(int value);
```

*In parameters*

value
    Non-negative integer value.

*Remarks*

    Important: The user should not use this function.

## Step-by-step instruction

This section contains step-by-step instructions on how to use the BSF-skeleton to quickly create a parallel program. Starting from Step 2, we strongly recommend compiling the program after adding each language construction.

**Step 1.** First of all, we must represent our algorithm in the form of operations on lists using the higher-order functions *Map* and *Reduce* (see Algorithm 1. Generic BSF-algorithm template.). An example is presented in Section "Example of using the BSF-skeleton".

**Step 2.** In the file *Problem-Parameters.h*, define problem parameters. For example:

```
#define PP_N 3 // Dimension of space
```

**Step 3.** In the file *Problem-Types.h*, declare problem types (optional). For example:

```
typedef PT_point_T[PP_N]; // Point in n-Dimensional Space
```

**Step 4.** In the file *Problem-bsfTypes.h*, implement the predefined BSF types. If we do not use a workflow then we do not have to implement the types *PT_bsf_reduceElem_T_1*, *PT_bsf_reduceElem_T_2*, *PT_bsf_reduceElem_T_3*, but we can't delete these empty structures. For example:

```
struct PT_bsf_parameter_T {
    PT_point_T approximation;        // Current approximation
};
struct PT_bsf_mapElem_T {
    int columnNo;                    // Column number in matrix Alpha
};
struct PT_bsf_reduceElem_T {
    double column[PP_N];             // Column of intermediate matrix
};
struct PT_bsf_reduceElem_T_1 { };
struct PT_bsf_reduceElem_T_2 { };
struct PT_bsf_reduceElem_T_3 { };
```

**Step 5.** In the file *Problem-Data.h*, define the problem-dependent variables and data structures. For example:

```
static double PD_A[PP_N][PP_N]; // Coefficients of equations
```

**Step 6.** In the file *Problem-bsfCode.cpp*, implement the predefined problem-dependent BSF functions (see Section "Predefined problem-dependent BSF functions (prefix PC_bsf_)") in the suggested order. To implement these functions, the user can write additional *problem (user) functions* in the *Problem-bsfCode.cpp*. The prototypes of these problem functions must be included in the *Problem-Forwards.h*.

**Step 7.** In the file *Problem-bsfCode.cpp*, we can configure the BSF-skeleton parameters (see Section "BSF-skeleton parameters").

**Step 8.** Build and run the solution in the MPI environment.

# Example of using the BSF-skeleton

In this section, we show how to use the BSF-skeleton to implement the iterative Jacobi method as an example. The *Jacobi method* [8] is a simple iterative method for solving a system of linear equations. Let us give a brief description of the Jacobi method. Let a joint square system of linear equations in a matrix form be given in Euclidean space $\mathbb{R}^n$:

$$Ax = b, \quad (1)$$

where

$$A = \begin{pmatrix} a_{11} & \cdots & a_{1n} \\ \vdots & \ddots & \vdots \\ a_{n1} & \cdots & a_{nn} \end{pmatrix},$$

$$x = (x_1, \ldots, x_n),$$

$$b = (b_1, \ldots, b_n).$$

It is assumed that $a_{ii} \neq 0$ for all $i = 1, \ldots, n$. Let us define the matrix

$$C = \begin{pmatrix} c_{11} & \cdots & c_{1n} \\ \vdots & \ddots & \vdots \\ c_{n1} & \cdots & c_{nn} \end{pmatrix},$$

in the following way:

$$c_{ij} = \begin{cases} -\dfrac{a_{ij}}{a_{ii}}, \forall j \neq i; \\ 0, \forall j = i. \end{cases}$$

Let us define the vector $d = (d_1, \ldots, d_n)$ as follows: $d_i = b_i/a_{ii}$. The Jacobi method of finding an approximate solution of system (1) consists of the following steps:

Step 1. $k \coloneqq 0; x^{(0)} \coloneqq d$.
Step 2. $x^{(k+1)} \coloneqq Cx^{(k)} + d$.
Step 3. If $\left\| x^{(k+1)} - x^{(k)} \right\|^2 < \varepsilon$, go to Step 5.
Step 4. $k \coloneqq k + 1$; go to Step 2.
Step 5. Stop.

In the Jacobi method, an arbitrary vector $x^{(0)}$ can be taken as the initial approximation. In Step 1, the initial approximation $x^{(0)}$ is assigned by the vector $d$. In Step 3, the Euclidean norm $\|\cdot\|$ is used in the termination criteria. The *diagonal dominance* of the matrix $A$ is a sufficient condition for the convergence of the Jacobi method:

$$|a_{ii}| \geq \left( \sum_{j=1}^{n} |a_{ij}| \right) - |a_{ii}|$$

for all $i = 1, \ldots, n$, and at least one inequality is strict. In this case, the system (1) has a unique solution for any right-hand side.

Let us represent the Jacobi method in the form of algorithm on lists. Let $c_j$ denotes the $j$-th column of matrix $C$:

$$c_j = \begin{pmatrix} c_{1j} \\ \vdots \\ c_{nj} \end{pmatrix}.$$

Let $G = [1, \ldots, n]$ be the list of natural numbers from 1 to $n$. For any vector $x = (x_1, \ldots, x_n) \in \mathbb{R}^n$, let us define the function $F_x: \{1, \ldots, n\} \to \mathbb{R}^n$ as follows:

$$F_x(j) = \begin{pmatrix} x_j c_{1j} \\ \vdots \\ x_j c_{nj} \end{pmatrix},$$

i.e. the function $F_x(j)$ multiplies the $j$-th column of the matrix $C$ by the $j$-th coordinate of the vector $x$. The BSF-implementation of the Jacobi method presented as Algorithm 3 can be easily obtained from the generic BSF-algorithm template (see Algorithm 1). In the algorithm 2, $\vec{+}$ and $\vec{-}$ denote the operations of vector addition and subtraction, respectively. Note that the matrix $C$ entered in line 1 is implicitly used to calculate the values of the function $F_{x^{(k)}}$ in line 3.

The source code of the BSF-Jacobi algorithm, implemented by using the BSF-skeleton, is freely available on Github at https://github.com/leonid-sokolinsky/BSF-Jacobi. Additional examples of using the BSF-skeleton can be found on GitHub at the following links:

- https://github.com/leonid-sokolinsky/BSF-LPP-Generator;
- https://github.com/leonid-sokolinsky/BSF-LPP-Validator;
- https://github.com/leonid-sokolinsky/BSF-gravity;
- https://github.com/leonid-sokolinsky/BSF-Cimmino;
- https://github.com/leonid-sokolinsky/NSLP-Quest.

# Workflow support

The BSF-skeleton supports workflows. A workflow consists of orchestrated and repeatable activities (jobs). The BSF-skeleton supports up to four different jobs. The starting job is always numbered 0 (omitted in the source codes). The other jobs have sequential numbers 1, ..., 3. Each job has its own type of reduce-list elements defined in the file *Problem-bsfTypes.h*. All jobs have the same type of map list elements. To organize the workflow, we need to follow these steps:

| | **Algorithm 3.** BSF-Jacobi algorithm with *Map* and *Reduce*. |
|---|---|
| 1: | **input** $C, d$ |
| 2: | $k := 0; x^{(0)} := d; G := [1, \ldots, n]$ |
| 3: | $B := Map(F_{x^{(k)}}, G)$ |
| 4: | $s := Reduce(\vec{+}, B)$ |
| 5: | $x^{(k+1)} := s \vec{+} d$ |
| 6: | $k := k + 1$ |
| 7: | **if** $\left\| x^{(k)} \vec{-} x^{(k-1)} \right\|^2 < \varepsilon$ **goto** 9 |
| 8: | **goto** 3 |
| 9: | **output** $x^{(i)}$ |
| 10: | **stop** |

1. In the file *Problem-bsfParameters.h*, redefine the macros *PP_BSF_MAX_JOB_CASE* specifying the largest number of a job. For example, if the total job quantity is 3, the number to be assigned to *PP_BSF_MAX_JOB_CASE* must be 2.
2. In the file *Problem-bsfTypes.h*, define the types of reduce-list elements for all jobs whose sequential numbers are less than or equal to *PP_BSF_MAX_JOB_CASE*.
3. In the file *Problem-bsfCode.cpp*, implement the functions *PC_bsf_MapF[_*]*, *PC_bsf_ReduceF[_*]*, *PC_bsf_ProcessResults[_*]*, *PC_bsf_ProblemOutput[_*]* and *PC_bsf_IterOutput[_*]* for all jobs whose sequential numbers are less than or equal to *PP_BSF_MAX_JOB_CASE*. The functions *PC_bsf_ProblemOutput[_*]* should assign the parameter **nextJob* a sequential number of the next job (possibly the same).

If the number of workflow states is greater than the number of jobs, you can use the function *PC_bsd_JobDispatcher* to manage these states. An example of a solution using the BSF-skeleton with the workflow support is freely available on Github at https://github.com/leonid-sokolinsky/Apex-method [9].

## OpenMP support

The BSF-skeleton supports a parallelization of the map-list processing cycle in the worker processes (the function *BC_WorkerMap*) using the `#pragma omp parallel for`. This support is disabled by default. To enable this support, we must define the macros *PP_SF_OMP* in the file *Problem-bsfParameters.h*. Using the macros *PP_BSF_NUM_THREADS*, we can specify the number of threads to use in the `parallel for`. By default, all available threads are used.

## Using Map without Reduce

Some numerical algorithms can be implemented naturally using the function *Map* without the function *Reduce* [10]. In this section, we will show how to use the BSF-skeleton in this case. As an example, we use the Jacobi method described above. Let $G = [1, \ldots, n]$ be the list of natural numbers from 1 to *n*. For any vector $x = (x_1, \ldots, x_n) \in \mathbb{R}^n$, let us define the function $\Phi_x: \{1, \ldots, n\} \to \mathbb{R}$ as follows:

$$\Phi_x(i) = d_i + \sum_{j=1}^{n} c_{ij} x_j, \qquad (2)$$

i.e. the function $\Phi_x(i)$ calculates the *i*-th coordinate of the next approximation. An implementation of the Jacobi method that uses only a higher-order function *Map* is shown in Algorithm 4. In this case, the reduce-list consists of coordinates of the next approximation and does not require performing *Reduce*. An implementation of Algorithm 4 using the BSF-skeleton is freely available on Github at https://github.com/leonid-sokolinsky/BSF-Jacobi-Map. In the implementation of the function *PC_bsf_MapF*, we had to apply a couple of tricks that use the skeleton variables *BSF_sv_numberInSublist*, *BSF_sv_addressOffset* and *BSF_sv_sublistLength* (see Section "Skeleton variables").

| | **Algorithm 4.** BSF-Jacobi algorithm with *Map*. |
|---|---|
| 1: | **input** $C, d$ |
| 2: | $k := 0; x^{(0)} := d; G := [1, \ldots, n]$ |
| 3: | $x^{(k+1)} := Map(\Phi_{x^{(k)}}, G)$ |
| 4: | $k := k + 1$ |
| 5: | **if** $\left\| x^{(k)} \vec{-} x^{(k-1)} \right\|^2 < \varepsilon$ **goto** 7 |
| 6: | **goto** 3 |
| 7: | **output** $x^{(i)}$ |
| 8: | **stop** |

## Supplementary material

The source code of the BSF-skeleton is freely available on Github at https://github.com/leonid-sokolinsky/BSF-skeleton.

## Declaration of competing Interest

The authors declare that they have no known competing financial interests or personal relationships that could have appeared to influence the work reported in this paper.

## Acknowledgements

This work was supported by the Russian Foundation for Basic Research [project No. 20-07-00092-a] and the Ministry of Science and Higher Education of the Russian Federation [gov. order FENU-2020-0022].